\documentstyle[prl,aps,epsfig,multicol]{revtex}
\begin{document}
\draft

\title{Influence of shear flow on vesicles near a wall: 
a numerical study}
\author{Sreejith Sukumaran\footnote{Electronic address : 
sreesuku@mpikg-golm.mpg.de} and Udo Seifert\footnote{Electronic
address : useifert@mpikg-golm.mpg.de}}
\address{Max-Planck-Institut f\"{u}r Kolloid- 
und Grenzfl\"{a}chenforschung, 
Am M\"{u}hlenberg 2, 14476 Golm, Germany}
\date{\today}
\maketitle

\begin{abstract}
We describe the dynamics of three--dimensional fluid vesicles 
in steady shear flow
in the vicinity of a wall.
This is analyzed numerically at low Reynolds numbers using a 
boundary element method.
The area--incompressible vesicle exhibits bending elasticity. 
Forces due to adhesion or gravity oppose the hydrodynamic 
lift force driving the
vesicle away from a wall. 
We investigate three cases. First, a neutrally buoyant vesicle 
is placed in the vicinity
of a wall which acts only as a geometrical constraint.
We find that the lift velocity is linearly proportional to shear 
rate and decreases with increasing
distance between the vesicle and the wall.
Second, with a vesicle filled with a denser fluid, we find a 
stationary hovering state.
We present an estimate of the viscous lift force 
which seems to agree with recent experiments of 
Lorz {\em et al.} [Europhys. Lett. {\bf 51}, 468 (2000)].
Third, if the wall exerts an additional adhesive force, we 
investigate the 
dynamical unbinding transition which occurs at an adhesion 
strength linearly proportional
to the shear rate. 
\end{abstract}


\begin{multicols}{2}

\section{Introduction}

We would like to understand how hydrodynamic flow affects 
vesicles and, in
particular, shear flow because it occurs whenever a fluid 
flows near a surface. 
This is of relevance in modeling artificial or natural cells 
in motion
from the level of the dynamics of a single cell to the 
rheology of a suspension
such as the blood.
It will also enable us to understand
the interplay between vesicles in shear flow and specific and 
nonspecific 
adhesive properties of a membrane. 
This is a key feature in phenomena such as leukocyte locomotion
\cite{bongrand,bruinsma}. 
Computational methods are now increasingly used to understand 
and analyze
experiments studying the relationship between receptor--ligand 
functional
properties and the dynamics of adhesion. 
An efficient numerical method simulating adhering spherical cells 
in flow has been used recently
to investigate the
physical kinetics of adhesion molecules \cite{hammer}. 
There are compelling reasons to go beyond spherical cells, {\em viz.} 
($a$) to understand how deformation  
couples with the dynamics of adhesion centers
and ($b$) to study the interplay between the physical kinetics of 
adhesion molecules,  
orientation of a cell and viscous lift forces that act on 
non--spherical cells and do not act on
spherical cells.
It should however be noted that leukocyte locomotion is further 
complicated 
by the role of actin polymerization, ion channels and pseudopod 
formation 
which together play an important role in leukocyte response to shear 
stress \cite{schonbein}.
Another practical use of this study concerns mechanosensory transduction 
which is the
process by which certain cells convert fluid stresses
to biochemical and/or electrical signals \cite{davies}.
The identification and activation of mechanoreceptors and intracellular 
signaling pathways are currently 
active areas of research. 
For an example of a case where our study is relevant, consider 
an experiment 
which investigates the influence of mechanical stress
on the activation of specific proteins present on the membrane of 
endothelial cells. One method
involves reconstituting the proteins into phospholipid vesicles which 
are then subjected to 
physiological levels of fluid shear stress in a viscometer \cite{frangos}.
Quantitative predictions
require information regarding hydrodynamic stresses and tensions
that develop in response to flow and the model discussed in this paper is 
suited for these purposes.

We are also motivated by recent experiments on weakly adhering vesicles 
in shear flow \cite{sackmann}. 
These investigations reveal the lipid flow within the 
vesicle membrane and the flow field within the vesicle and near its 
outer surface.
The influence of weak flow fields on the state
of adhesion and the translational motion of vesicles are studied using 
bright
field and reflection interference contrast microscopy. Moreover, the 
hydrodynamic
lift forces exerted on adhering vesicles are estimated and the lift 
forces are found 
significantly larger than estimates based on an earlier 
theoretical study \cite{bruinsma}. An explanation for this discrepancy
based on arguments using a lubrication analysis of flow between a spherical 
cap and a wall has been proposed recently \cite{udo}.

Here, we show the results of a numerical simulation. 
With the above experiment in mind, we simulate an incompressible
vesicle in shear flow which is influenced by gravitational forces 
or nonspecific 
adhesion forces to be near a wall and by viscous lift forces to be
away from the wall. The main objectives of this paper
are ($a$) to understand the very nature 
of viscous lift force; ($b$) to describe how the steady state shape 
is influenced by shear flow and forces due to gravity or an adhesion 
potential;  
and ($c$) to study the influence of shear in taking the system 
away from equilibrium. The numerical method simulates a realistic 
model of a vesicle. The tensions that develop in response to flow
are determined 
self--consistently.

Before describing the study, we take stock of what is known about 
the dynamics of vesicles, and
about techniques that are relevant. 
The dynamics of vesicles 
belong to a broad class of problems called free--interface problems 
which describe 
the dynamics of a particle consisting of a membrane that encloses a 
drop of an
incompressible Newtonian fluid often
called a capsule in the literature \cite{barthes}. The first problem 
discussed in this class
was the behaviour of a fluid drop in shear flow,
when the effects of fluid viscosity and interfacial tension are 
taken into  
account \cite{taylor}. 
In fact, much of the later work on elastic capsules or vesicles employed 
techniques and analysis used to understand the dynamics of liquid drops. 
The first two lines of attack involved analytical methods based on 
perturbation
schemes for shapes close to a sphere, and theories using slender--body 
dynamics.
Considerable effort has been made in fluid drop dynamics ranging from
drop breakup to physical influences such as surfactants and complex flows,
for reviews see \cite{rallison,stone}. An important step in 
understanding this
nonlinear problem of three--dimensional deformation in external flows 
without 
making drastic approximations involves numerical methods 
using boundary element methods \cite{pozrikidis}. 
The use of an unstructured grid of triangular elements to represent the
interface has been shown to be efficient and numerically stable 
in simulating 
the dynamics of three--dimensional liquid droplets in shear flow 
in an unbounded fluid and in the proximity of a bounding 
plane wall \cite{pozrikidis1}.
This has been also used in simulating capsules with elastic
membranes \cite{pozrikidis2}.
We now compare this scene of activity with that regarding the dynamics
of vesicles. 

In the last three decades, experimental and theoretical studies of 
the behaviour of free and bound 
fluid vesicles in equilibrium have met with great success \cite{seifert}.
But, the hydrodynamics of vesicles still poses unsolved problems.  
For example, consider a vesicle in unbounded shear flow. 
A minimal model of a vesicle involves bending 
elasticity and the constraints of fixed area and volume. 
Using this model, the dynamics of free vesicles
in shear flow have been studied using boundary element
methods \cite{kraus,kraus1}.
Apart from features similar to liquid drops such as steady revolution 
of the surface, called 
tank--treading in the case of capsules, and 
a steady tilt, the numerical work suggested that shear is a 
singular perturbation
even though one would naively expect that at very small shear rates 
the equilibrium shape 
is recovered. Numerical limitation prevented a conclusive examination
of this regime.

In {\em two--dimensions}, the dynamics of adhering vesicles 
has been investigated in models for chemotaxis \cite{misbah1}
and in shear flow \cite{misbah} using a boundary element method. 
Practical considerations and a need to 
compare theoretical models with experiments call for an investigation 
of how hydrodynamic flows
affect {\em three--dimensional} vesicles.

In the next section, the physical model of a vesicle and the 
mathematical formulation 
of the hydrodynamics of vesicles including the implementation of the 
boundary element method will be briefly discussed.
Then, we proceed to three sections concerning the dynamics of vesicles 
in shear flow in the proximity
of a wall. In Section III, we study the nature of viscous hydrodynamic 
lift force and
we assume that the vesicle does not experience gravitational 
force, {\em i.e.}, the vesicle is neutrally
buoyant, and the vesicle is not influenced by forces due 
to an adhesion potential. 
In Section IV, gravitational forces are introduced and we compare 
our results with
the experiment. In Section V, the influence of a nonspecific 
adhesion potential
is considered. And in Section VI, our conclusions regarding 
the dynamics of
vesicles near a wall are summarized.

\section{Model}

Consider the vesicle to be a two--dimensional surface embedded 
in three--dimensional
space. The instantaneous membrane configuration $ {\bf R}(s_1,s_2) $ is 
parametrized by internal coordinates $(s_1,s_2)$.
The energy, {\em E}, of the vesicle -- with an area element $dS$ --

\begin{equation}
E \equiv \oint dS \left[ \frac{\kappa}{2} (2 H)^2
+ \Sigma + \cal{W} \right],
\end{equation}

\noindent has three contributions. The first due to the squared mean 
curvature, $ H^2 $, describes 
the bending energy \cite{seifert} with bending rigidity $\kappa$.
The second term is due to a locally varying isotropic tension $\Sigma$ 
which is needed to ensure
local incompressibility of the membrane. The third term is 
the adhesion energy 
due to the proximity of a homogeneous substrate which exerts 
a nonspecific 
interaction. The geometry is schematically represented in Figure 1.
The adhesion potential, with the wall in the plane $ z = 0 $, is
chosen to be

\begin{equation}  
{\cal W}(z) \equiv W (d_0/z)^2 \left[ (d_0/z)^2 - 2 \right].
\end{equation}

\noindent Here, $W$ is the adhesion strength. 
The potential is repulsive as $1/z^4$ for $z \ll d_0$ and attractive as
$-1/z^2$ at long distance. The potential has a 
minimum of $ - W $ at $ z = d_0 $.
In the case of free vesicles, the adhesion term is absent. 
The membrane force density, $ {\bf f} $, reads 

\begin{equation}
{\bf f} ({\bf R}) \equiv - \left(
\frac{1}{\sqrt{g}} \frac{\delta E}{\delta {\bf R}} \right),
\end{equation}
\noindent where $g$ is the determinant of the metric tensor.

To describe the hydrodynamics of vesicles, we assume that the 
Reynolds number of the flow inside and around the vesicle based upon the 
vesicle size (say the radius of the spherical vesicle of equal volume) is
sufficiently small so that the velocity, ${\bf v}$, of the fluid
and the pressure field, $ p $, is governed
by the equations of Stokes flow or creeping flow equations. Thus

\begin{eqnarray}
\eta^{out} \partial_{jj} v^{out}_i = \partial_i p^{out} & , 
\hspace{1cm} & \partial_i v^{out}_i = 0, \nonumber \\
\eta^{in} \partial_{jj} v^{in}_i = \partial_i p^{in} & , 
\hspace{1cm} & \partial_i v^{in}_i = 0, 
\end{eqnarray}

\noindent where $p^{out}$ and $p^{in}$ are the pressures, 
and $\eta^{out}$ and $\eta^{in}$ 
are the viscosities associated with the outer and inner fluids 
respectively, and $ \partial_i $ denotes differentiation with 
respect to the coordinate $x_i = x,y,z $ for $ i=1,2,3$ respectively.
The summation convention is used over 
doubly occurring indices. 

If the vesicle is not neutrally buoyant, then we account for the
body force acting on the vesicle due to gravity. This is done by 
modifying the membrane force density. 
The modified force density, $ {\bf f}^{mod} $, is given 
by \cite{pozrikidis}

\begin{equation}
{\bf f}^{mod}({\bf R}) = {\bf f}({\bf R}) + (\rho^{in} - 
\rho^{out}) \hspace{0.1cm} 
({\bf g} \cdot 
{\bf R}) \hspace{0.1cm} {\bf n}({\bf R}).
\end{equation}

\noindent 
Here, $\rho^{in}$ and $\rho^{out}$ are the densities of the outer 
and inner fluids
respectively; ${\bf n}$ is the unit vector normal to the vesicle, 
${\bf g} \equiv -g_0 {\hat {\bf e}}_z $ is the acceleration due 
to gravity where
$ g_0 \simeq 9.81 $ m s$^{-2}$  and $ {\hat {\bf e}}_z $ is the 
unit vector in the $z$-direction.

To express the no--slip
condition on bounding surfaces and also the impermeability of the 
membrane, we require that the velocity be continuous across the 
vesicle.
This also provides the kinematic condition by which the vesicle 
shape changes
with time.
The hydrodynamic surface force is allowed to undergo a discontinuity 
that is balanced by the membrane forces, $ {\bf f}^{mod}({\bf R}) $.

Rather than solve for the fluid velocity at all points in space, it is 
advantageous to use a boundary--integral method by which the Stokes 
equations 
inside and outside are cast into an integral form that involves only 
quantities evaluated on the vesicle surface \cite{pozrikidis}. 
This formalism provides us
with an integral equation for the membrane velocity in 
terms of $ {\bf f} $,

\begin{equation}
v_i({\bf R}^{\alpha}) = v_i^{\infty}({\bf R}^{\alpha}) + 
\frac{1}{8 \pi \eta }
\oint G_{ij}({\bf R}^{\alpha},{\bf R}^{\beta}) f_j({\bf R}^{\beta}) dS,
\end{equation}

\noindent where $ {\bf v}^{\infty} (x,y,z) = (\dot{\gamma} z, 0, 0) $ 
is the incident shear flow 
(refer Figure 1) and
$ G_{ij}({\bf R}^{\alpha},{\bf R}^{\beta}) $ is the appropriate
Green's function for the velocity. In the case of free vesicles 
in unbounded space,
the free--space Green's function is the Stokeslet; and, in the 
case of bound vesicles, we choose the semi--infinite space Green's 
function for the 
velocity \cite{note1}. 
To make computations tractable, we assume that 
the viscosity of the fluid inside the vesicle is equal to that of 
the suspending fluid,
$ (\eta^{out} = \eta^{in} = \eta) $.
If the viscosities are different, equation (6) is modified to 
include an additional term 
which turns it into an integral equation in {\bf v}. Then, the 
equation has to be solved
iteratively or {\bf v} must be determined by matrix inversion at 
any instant of time. 
We also note in passing that by choosing the same viscosity,
we do not allow for tumbling motion in shear flow which occurs   
when the dynamics is vorticity--dominated. For liquid drops, tumbling 
is shown to occur for $ \lambda \equiv \eta^{in}/\eta^{out} \geq 
4 $ \cite{pozrikidis1}.

We characterize vesicles by a dimensionless number, the reduced volume 
\begin{equation}
v \equiv V/(4 \pi R_0^3/3) ,
\end{equation} 
where $V$ is the enclosed volume and the surface area $A$
determines the length scale $ R_0 = \sqrt{A/(4 \pi)} $. For a sphere, $v=1$.
The bending rigidity of the 
membrane $\kappa$ sets the energy scale. We do not incorporate thermal 
fluctuations.
Throughout this paper, unless otherwise mentioned, lengths are expressed
in units of $R_0$,
the adhesion strength in units of $ \kappa/R_0^2 $, and the
shear rate in units of $ \kappa/(8 \pi \eta R_0^3) $. When gravitational 
forces are 
considered, the dimensionless gravity parameter 
\begin{equation}
g' \equiv (\rho^{in} - \rho^{out}) g_0 R_0^4/\kappa 
\end{equation}
measures this effect.
In simulations involving the adhesion potential equation (2), we 
take $d_0 = 0.01 R_0 $.

The implementation of the boundary element method is now 
well--established \cite{pozrikidis1,pozrikidis2,kraus,kraus1}. Here, we 
briefly mention 
the essential steps. \\
\indent ($a$) The surface is 
approximated by a grid of triangular elements. We take 512 triangles 
with 258 nodes. \\
\indent ($b$) The force density is calculated at the nodes. \\
\indent ($c$) Along with the condition dictated by 2D incompressibility,
\begin{equation}
\frac{\partial}{\partial t} \sqrt{g} = \partial_i v_i - n_i n_j 
\partial_j v_i = 0 ,
\end{equation} 
the integral equation equation (6)
yield equations for the local tensions $\Sigma $. 
Unlike droplets with constant tension, the tensions $\Sigma $ 
develop so that 
the membrane element deforms while maintaining the original area. \\
\indent ($d$) After computing $\Sigma $, the integral equation 
(6) is solved for ${\bf v}$. \\
\indent ($e$) The position of the nodes are advanced, the 
coordinates updated
and we return to step ($b$) unless a steady state is reached.

Before tackling cases involving externally imposed 
hydrodynamic flows, we test if
the numerical scheme yields the stationary shapes that are known to
exist in equilibrium conditions, {\em i.e.}, neutrally buoyant 
free vesicles \cite{seifert}.  
The minimal model we have chosen allows a wide spectrum of shapes. 
For $ 0.65 \lesssim v \lesssim 1 $, the stable states are 
prolate--like spheroids. 
For $ v \lesssim 0.75 $, oblate discocytes are locally stable. 
So far, there is no indication of the presence of locally 
stable non--axisymmetric shapes in this model.
For $ {\bf v}^{\infty} = {\bf 0} $, our numerical scheme 
gives the stable and 
metastable axisymmetric shapes, including dumb--bell shaped 
prolates and erythrocyte--like 
biconcave discocytes.  
In Figure 2, we show an example in which the initial oblate spheroid 
relaxes to the final equilibrium
(stationary) biconcave
shape. 
Within this minimal model, the vesicle relaxes to the next 
dynamically accessible
locally stable shape. Depending 
on the initial condition, this is either a prolate, an oblate
or a stomatocyte.

In contrast to vesicles, 
the shape of {\em incompressible} interfaces that are not influenced 
by curvature elasticity
is infinitely degenerate.
In such cases, the stationary shape depends only upon the membrane 
surface area, and as a 
consequence of neglecting elastic behaviour, the interface 
can retain any 
shape with the same surface area.
At the other end of the spectrum, liquid drops with interfaces  
influenced by a constant surface tension and also compressible 
interfaces which are only influenced by
curvature elasticity remain spherical. 

\section{Unbound neutrally buoyant vesicles near a wall}

We now simulate the influence of steady shear flow on 
vesicles in the
proximity of a wall.
To retrieve the bare contribution of hydrodynamics to the lift force,
adhesion and gravity are not considered
in this section. The vesicle is
initially at a small distance from the wall and 
then, shear flow is imposed.  
In Figure 3, a typical sequence of ``snapshots'' are shown. 
In the far--field, the vesicle translates with a speed of the 
same order of the difference of velocity,
due to external shear flow, that exists across the vesicle. The 
vesicle also
experiences a lift velocity. From Figure 3, it can be extracted 
that the 
lift velocity is an order of magnitude smaller than the 
translational velocity.

Analyzing several cases, we find the main characteristics in this 
setup to be: \\
\indent ($i$) the membrane of the vesicle tank--treads; \\
\indent ($ii$) the vesicle develops a steady tilt which is roughly 
independent of the shear rate; and  \\
\indent ($iii$) the shape changes to a prolate--like ellipsoid;  \\
This is similar to the
dynamics of a free vesicle in unbounded shear flow \cite{kraus}.
The first two features are understood \cite{note4}.
The tank--treading motion is due to the rotational component of 
linear shear flow.
The balance of the
moments on the vesicle due to the effect of shear-flow acting on
a stationary inclined vesicle 
and that due to the tank--treading motion results in a steady 
tilt. These two moments are linearly 
proportional to shear rate and hence, the tilt is independent of 
the shear rate in the steady state. In 
the case of
liquid droplets, the tilt reduces more drastically with shear rate 
because of the elongation 
of the droplet, and thereby a reduction in $v$. But in the case of 
vesicles considered 
here, $v$ is a constant due to incompressibility. As mentioned in 
the introduction,
for the third feature regarding 
the prolate ellipsoidal shape, it is still not 
understood whether it will persist for arbitrarily small shear rates.  

From the above discussion, it is clear that the steady tilt of the 
vesicle plays 
the leading role in breaking 
the fore--aft symmetry of the vesicle with respect to shear flow. Since
the tilt
is independent of shear rate, the excess pressure in the space between 
the vesicle and the wall
can be expected to cause a lift force which is
proportional to shear rate, $ \dot{\gamma} $. 
In fact, the dynamics of a vesicle near a non--adhering wall
is roughly similar for any shear rate; and as a
function of $ \tau \equiv \dot{\gamma} t $, the ``snapshots'' as shown 
in Figure 3 will be similar
for vesicles with the same reduced volume.
In Figure 4, we plot for different shear rates 
the ratio of lift velocity, $v_{lift}$, to shear rate against the mean 
height.
The data collapse indicates that $ v_{lift} \propto \dot{\gamma} $.  

We now derive an empirical expression for the lift velocity guided by 
the far--field 
result obtained for
ellipsoidal cells:
$ v_{lift} = U \dot{\gamma} (R_0^3/h^2)  $
where $ U \simeq 0.1-0.3 $ for $ v \simeq 0.9-0.99 $ \cite{olla}.  Assuming 
the same power 
law, we get
an estimated asymptotic fit for the lift velocity to be
\begin{eqnarray}
v_{lift} & \simeq & 0.08 \hspace{0.1cm} \dot{\gamma} \frac{R_0^3}{h^2} 
\hspace{1cm} (v=0.95) 
\end{eqnarray}
Here, $h$ is the mean distance of the center of the vesicle to the 
substrate.
We have simply calculated the mean center and the mean velocity to be 
the arithmetical mean of the
coordinates and velocity of the nodes respectively. Then, $v_{lift}$ is
just the vertical
or $z$-component of the mean velocity. The fit needs to be taken 
with caution due to three reasons.
First, the exact dependence of $ v_{lift} $ on $ h $ is not easy to 
access with the limited 
range as shown in Figure 4. 
Second, the asymptotic fit 
(and also, far--field results based on analytical methods) for
vesicle migration assumes that the problem can be treated 
in a quasistatic way. 
This implies that it is strictly valid only when the time--scale 
for deformation
and shape changes is much smaller than the time--scale for migration.
Since our numerical method is similar to an experimental situation 
and we do not
arbitrarily fix the position of a vesicle, the fit is probably 
valid only for
large shear rates. This is partially evident from Figure 4 where 
we see that the 
plots for lower shear rates do not exactly collapse onto the plots 
for higher shear rates. Third, 
there is no reason to expect asymptotic results to hold good 
for $ h \simeq 0.25-0.5 R_0 $. But,
there is also no reason to be alarmed if it is true in that range. 
In fact, analogous results for liquid drops do indeed match remarkably well 
for $ \lambda = 1 $ at $ h \simeq 0.25-0.5 R_0 $ \cite{pozrikidis2}.
We can also infer from far--field results 
that the sign of the prefactor is determined by the sign of the 
tilt $ \theta $ of the vesicle.
Hence, when $ \dot{\gamma} \rightarrow -\dot{\gamma} $ and $ 
\theta \rightarrow - \theta $,
$ v_{lift} \rightarrow v_{lift} $ which agrees with expectations 
that the direction of migration 
should be independent of the direction of shear flow.

We emphasize that the dominant effect is due to the tilt of the vesicle 
and this leads to a lift force, $ F_{lift} \propto \dot{\gamma} $. 
Additional effects 
due to deformation can be expected to be of $ O(\dot{\gamma}^2) $. 

For $ h \simeq R_0 $, the asymptotic lift velocity 
can be taken to be the lower bound for the lift velocity. The actual
velocity could be expected to be higher due to transient effects or 
due the effect of ``touching'' the wall.  
Assuming Stokes' result $6 \pi \eta R_0 v_{lift} $,
this lower bound for the viscous lift force can be estimated from 
equation (10) for $v = 0.95$ to be
\begin{equation}
F_{lift} \simeq 0.45 \pi \eta \dot{\gamma} \frac{R_0^4}{h^2} 
\end{equation}
With this estimate of the lift force, a comparison
can be made with the results of the experiment. We shall do so only 
at the end of the next section,
after investigating the influence of gravity. 

\section{Influence of gravity}

The effect of gravity on vesicles arises from the frequently 
employed experimental
technique to stabilize the vesicle at the bottom of the measurement 
chamber by a difference in density
between the fluids inside and outside the vesicle. In the 
experiment \cite{sackmann},
this is done by choosing iso-osmolar but different buffers for 
swelling the vesicle
and suspending the vesicle. The threshold velocity at which 
unbinding of the vesicle from 
the substrate occurs allows to estimate the lift force. After 
unbinding, the vesicle 
hovers at a distance at which electrostatic forces or van der Waals forces 
that cause adhesion are negligible.
The lift force in the hovering state can therefore be assumed to be 
counteracted by gravitational forces only.

We simulate the influence of gravitational force on a vesicle. 
As in the previous section, we place the vesicle initially at a distance 
$h_0 \simeq 0.1 R_0 - 0.5 R_0$
and then, shear flow is applied. It would of course be more realistic 
if we started with a 
vesicle which rested on the wall, but at gravitational strengths 
which are of
relevance, this initial configuration leads to numerical instabilities.
Instead, the initial $h_0$ is allowed to be non-zero but we check if the  
steady state dynamics is independent of $h_0$ and the simulations  
confirm this view. We note in passing that the steady state 
depends on the reduced volume of the vesicle but does not depend 
on the initial shape.

As expected, gravity counteracts viscous lift force and the
vesicle hovers at a distance away from the wall. We now study how 
this distance depends 
on the shear rate. 
It is clear from Figure 5(a) that the initial $h_0$ can be chosen so 
that we can optimize 
the computational time in evolving the vesicle to the steady state.  
At large shear rates, the vesicle is sufficiently 
far from the wall and as discussed in the previous section, the lift is 
determined by the tilt alone. Meanwhile, at low shear rates, the 
vesicle is also deformed by 
the proximity to the wall. The dependence of the steady state mean height
on shear rate is shown in Figure 5(b) and seems to indicate those 
two regimes.

Let us now consider the three--dimensional shape of the vesicle in 
the steady state. 
In Figure 6, we show a comparison of the profile and also the top 
view of the vesicle
for different shear rates. We note the features that are immediately 
apparent from Figure 6. \\
\indent ($i$) Vesicles with small excess areas $ v \rightarrow 1 $ do 
not exhibit large
deformations with increasing shear rate. \\
\indent ($ii$) At large shear rates [refer Figures 6(c) and 6(d)], the 
shape
of the vesicle is roughly independent of shear rate. \\
\indent ($iii$) At any shear rate, the part of the vesicle away from 
the wall
shows roughly the same inclination. \\
\indent ($iv$) At lower shear rates [refer Figures 6(a) and 6(b)], 
gravity reduces the vesicle--wall distance. 
To allow for greater contact area, the shape changes from a prolate--like 
to an oblate--like
ellipsoid. \\
\indent ($v$) The tilt of the contact area decreases with decreasing  
shear rate. There is no unique definition of this tilt and therefore, the 
tilt of 
the ``base'' is difficult to quantify.

Now, we compare our results with the experiment \cite{sackmann}. There 
are some 
problems in making this comparison. The reduced volume $v$ and the 
gravitational strength $g'$
are not exactly known. For vesicles with very small excess area, the 
experiment does not
report any measurable tilt of the contact area nor the overall shape. Such 
a picture
fits in with our description only in the limit $g' \rightarrow \infty$. 
We take the shape of the vesicle to be close to a sphere with $ v = 0.95 $.
We assume the values quoted, {\em viz.} the density difference, 
$ \rho^{in} - \rho^{out} \simeq 5.2 $ mg/ml,
the radius of the vesicle $R_0 \simeq 14 \mu$m,
the vesicle--wall 
gap $ h_0 \simeq 100$nm
and shear stresses typically $ \eta \dot{\gamma} \simeq 1.2$mPa.
These experimental values lead to an estimated 
gravitational force on a vesicle with volume $V$ of
\begin{equation}
 F_{grav} = (\rho^{in} - \rho^{out}) g_0 V \simeq 6 \times 10^{-13} $N$.
\end{equation} 
\noindent As a theoretical estimate, we obtain
the lift force based on equation (11) assuming that $ h \simeq R_0 $ as 
\begin{equation}
 F_{lift} \simeq 0.45 \pi \eta \dot{\gamma} R_0^2 \simeq 3 
\times 10^{-13} $N$.
\end{equation} 
\noindent The agreement should be taken with a grain of salt if one 
takes into consideration 
all the assumptions that have gone into these estimates. 
 
\section{Influence of adhesion}
 
Now, we consider the dynamics of vesicles adhering to a surface 
due to nonspecific interactions. 
Here, the interplay between
forces due to an adhesion potential, equation (2),
and hydrodynamic forces
is considered. 
The vesicle is assumed to be neutrally buoyant.

Like an experiment, we start with an adhering vesicle, 
then apply shear flow and watch how the vesicle responds.
The dynamical evolution in the transient stage is shown in Figure 7.
These ``snapshots'' are qualitatively similar for any 
adhering vesicle with arbitrary reduced volume at any adhesion strength 
or shear rate.
The tilting of the vesicle is similar to the scenario in the 
previous section. 

The steady state of the adhering vesicle, of course, depends on the 
reduced volume, adhesion strength and shear rate.
Unlike the case involving gravity, vesicles adhering to a substrate
by the influence of a short--range potential cannot be made to hover at 
any arbitrary height by tuning
the strength of adhesion. 
There is a critical shear rate above which the vesicle 
unbinds from the wall.
For smaller shear rates, the vesicle tank--treads along the substrate.
In Figure 8, we show how the critical shear rate varies with 
increasing adhesion strength
for a vesicle with reduced volume, $v = 0.95 $. 
For $ \dot{\gamma} \lesssim \dot{\gamma}_c $, the steady state shape 
of the vesicle 
resembles the shape shown in Figure 7(c). Two main features of 
this state are to be noted. 
One, the tilt of the vesicle is nearly the same as that of 
a non--adhering vesicle near a wall,
and not very different from the tilt of a free vesicle in shear flow.
Two, the surface of adhesion is nearly zero. The vesicle can be 
said to be ``pinned''.
This state is nearly similar for any adhesion strength
and this explains the roughly linear relationship between the 
critical shear rate, $ \dot{\gamma}_c $, and
the adhesion strength.
For lower shear rates or higher adhesion strengths, the area of 
contact increases
and a typical example of a ``bound'' steady state is shown in Figure 9.
We find that the process of 
unbinding, {\em viz.} ``bound'' $\rightarrow$ ``pinned'' $\rightarrow $
unbound or ``free'', 
is similar to that described with simulations of the two--dimensional 
model \cite{misbah}.

Finally, we summarize the results of the simulations with a dynamical 
phase diagram.
First, we recall 
the behaviour of adhering vesicles under equilibrium conditions. 
In the absence of shear, it is known that there are two 
states \cite{seifert}.
For weak adhesion, the shape of the vesicle resembles the free shape 
and the surface
of adhesion is zero.
This state is called the pinned state. 
It should however be noted that there is no unique definition of 
the pinned state
in the dynamical picture.
When adhesion is strong, there is the bound state with non--zero 
adhesion area.
The adhesion area also depends on the excess area which is a measure 
of the reduced volume.
For $0.5 < v < 1$, the transition from the bound axisymmetric shape 
to the pinned state 
is discontinuous. 
Keeping in mind the similarity between the equilibrium and dynamical 
situations, we can 
construct a schematic dynamical phase diagram (refer Figure 10). 
Apart from the two--stage dynamical unbinding,   
we expect in the limit of weak adhesion, a regime in which the 
unbinding goes 
through only one stage. In the case of an adhesion potential with 
finite range,
there will be a finite critical shear rate for unbinding a vesicle 
which is pinned
under equilibrium conditions. If the vesicle is pinned by a contact 
potential, the vesicle
will unbind for any shear rate. This regime is difficult to realize 
either numerically 
or experimentally. 
In principle, there could be a direct transition between bound and 
free vesicles but 
we have not found any evidence for it
based on  
the numerical simulations over a wide range of parameters.

\section{Conclusion}

We conclude this study of the dynamics of vesicles near a wall 
with observations
regarding the computation, summary of main results, and certain 
remarks about 
current limitations and ongoing research. 
Using the boundary element method with a grid of triangular elements, 
we are able to simulate incompressible 
vesicles in bounded and unbounded shear flow. With moderate shear rates,
adhesion and gravitation strengths, the numerical method is stable 
and effective in studying the steady state of adhering, hovering or free
vesicles. The algorithm which is used to ensure area incompressibility
is the main hurdle in improving the speed of computation and also in using 
finer discretization. 

We are able to find the  
following features regarding the nature of viscous lift force on 
incompressible vesicles with bending elasticity: \\
\indent ($i$) The lift force due to the tilt of the vesicle is 
linearly proportional to the shear rate. It is the dominant part
in our simulations also for small shear rates. 
\\
\indent ($ii$) Our numerical results agree well with a 
recent experiment. \\ 
\indent ($iii$) The general features of dynamical unbinding is similar 
to that in
simulations of the two--dimensional model even though the additional 
possibility 
of a prolate--oblate shape change complicates the issue. 

We also remark about
interesting issues that need to be resolved, both experimentally 
and theoretically.
We have only considered a situation where the viscosities of the 
inside and outside fluid are equal. 
This is definitely an oversimplification of experiments with cells 
though it is not 
a limitation for experiments with phospholipid vesicles.
We still have to understand whether the model considered is singular
in the limit of very small shear rates.
We have 
not done simulations with the ratio of vesicle--wall gap to 
radius, $ h_0/R_0 < 0.001 $.
In this limit of vanishing gap, features such as the roughness 
of the wall
and also, the no--slip boundary condition will have to be reassessed. 
Finally, we note that the numerical method discussed here is 
amenable to an 
extension that includes specific receptor--ligand type of adhesion.

\end{multicols}

\eject

\begin{figure}
\begin{center}
\vspace{2cm}
\includegraphics[width=6in]{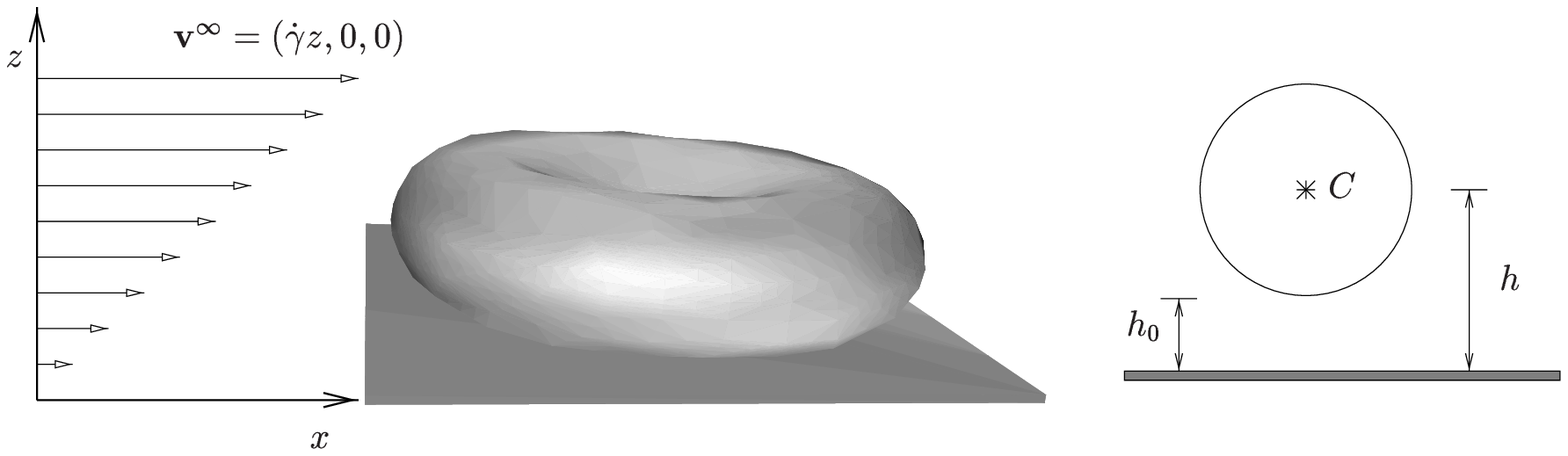}
\vspace{2cm}
\narrowtext
\caption{Schematic representation of the geometry showing a vesicle, 
the wall in the plane $z=0$ and applied linear shear flow. On extreme
right, the mean center $(x_C,0,z_C)$ is shown as $C$. The 
height of the center, $h$, and the vesicle--wall distance, $h_0$, are
also shown.}
\end{center}
\end{figure}

\eject

\begin{figure}
\begin{center}
\vspace{2cm}
\includegraphics{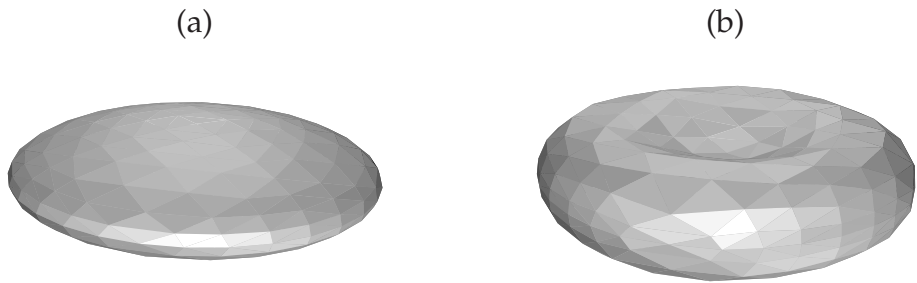}
\vspace{1cm}
\narrowtext
\caption{An example of a test of the numerical method in which 
an initial shape relaxes to
a stationary shape known to exist in the 
absence of externally applied flow.
(a) The initial shape is chosen to be an oblate convex spheroid 
with reduced volume, $ v = 0.6 $.
(b) After relaxation, the equilibrium or stationary shape is
a stable oblate biconcave discocyte with the initial area and volume. }
\end{center}
\end{figure}

\eject

\begin{figure}
\begin{center}
\vspace{2cm}
\includegraphics[width=7in]{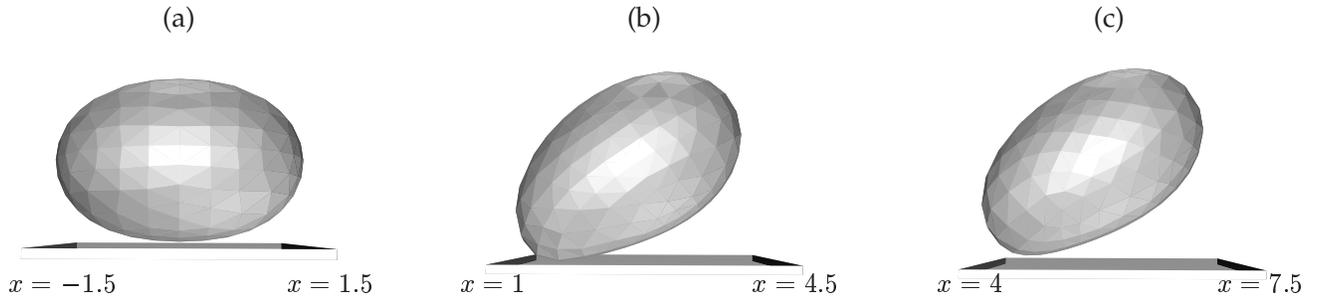}
\vspace{1cm}
\narrowtext
\caption{ ``Snapshots'' of the lift--off of an unbound vesicle away from 
a wall in the presence of shear flow. The 
plane $z=0$ shown is the wall. The $x$--coordinate
at the endpoints of the wall is shown at each instance of time, $t$. 
Here, $v = 0.95 , \dot{\gamma} = 30 $. 
(a) At $ t = 0 $, the initial shape is an oblate--like spheroid. 
The vesicle is at a distance $h_0 = 0.1 R_0$.
(b) The shape is now a prolate--like ellipsoid
and the vesicle is tilted with respect to the shear plane. 
Here, $t = 0.075$. 
(c) The shape and tilt of the vesicle is roughly the same as before,
and the lift--off is now clearly seen at $ t = 1.5 $. All quantities are
given in dimensionless units as discussed in the text.}
\end{center}
\end{figure}

\eject

\begin{figure}
\begin{center}
\vspace{2cm}
\includegraphics{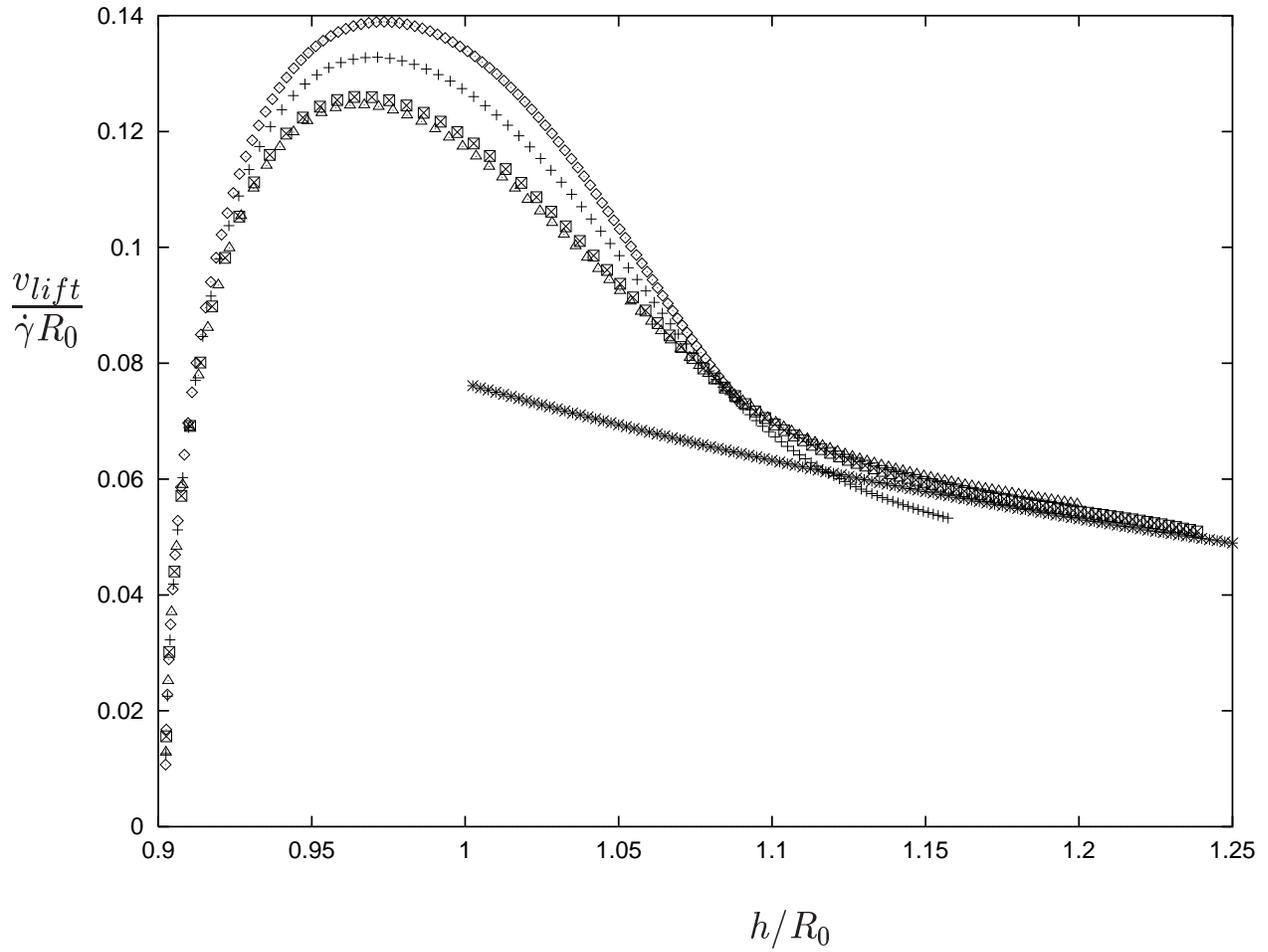}
\vspace{1cm}
\narrowtext
\caption{ The ratio of mean lift velocity, $v_{lift}$, to shear rate times 
the radius of vesicle, $\dot{\gamma} R_0$, 
is plotted against the ratio of the mean height $h$
of the vesicle from the wall to the the radius of vesicle for 
different shear rates
(definitions are given in the text). 
The lift velocity of the vesicle is roughly proportional to the shear rate.
Here, $v = 0.95 $, and in dimensionless units,
$ \dot{\gamma} = 6 (\diamond), 10 (+), 30 (\Box),
50 ( \times ) $ and the estimated asymptotic fit given by 
equation (10) is plotted using the asterisk ($\ast$).}
\end{center}
\end{figure}

\eject

\begin{figure}
\begin{center}
\vspace{2cm}
\includegraphics[width=4.5in]{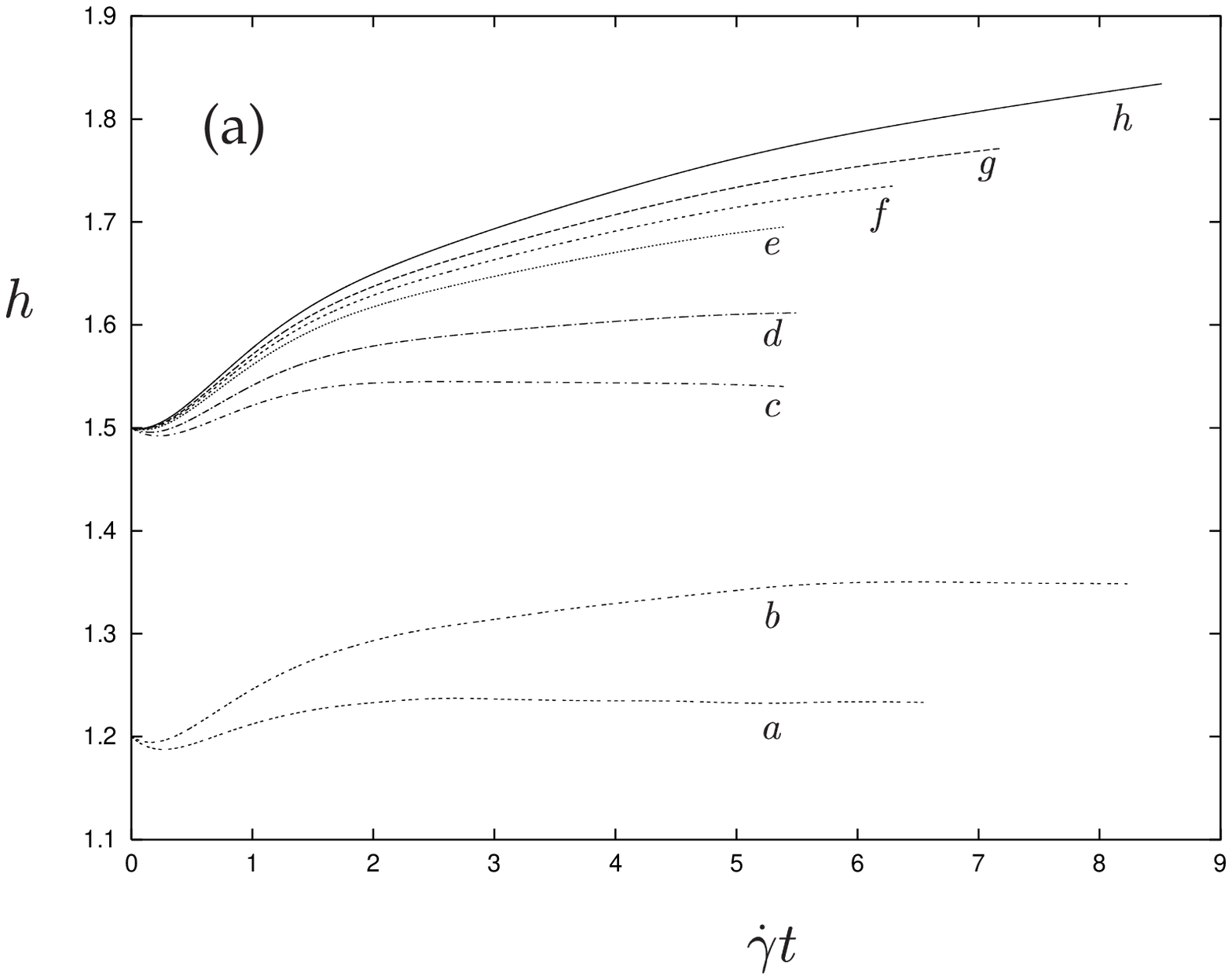} 
\includegraphics[width=4.5in]{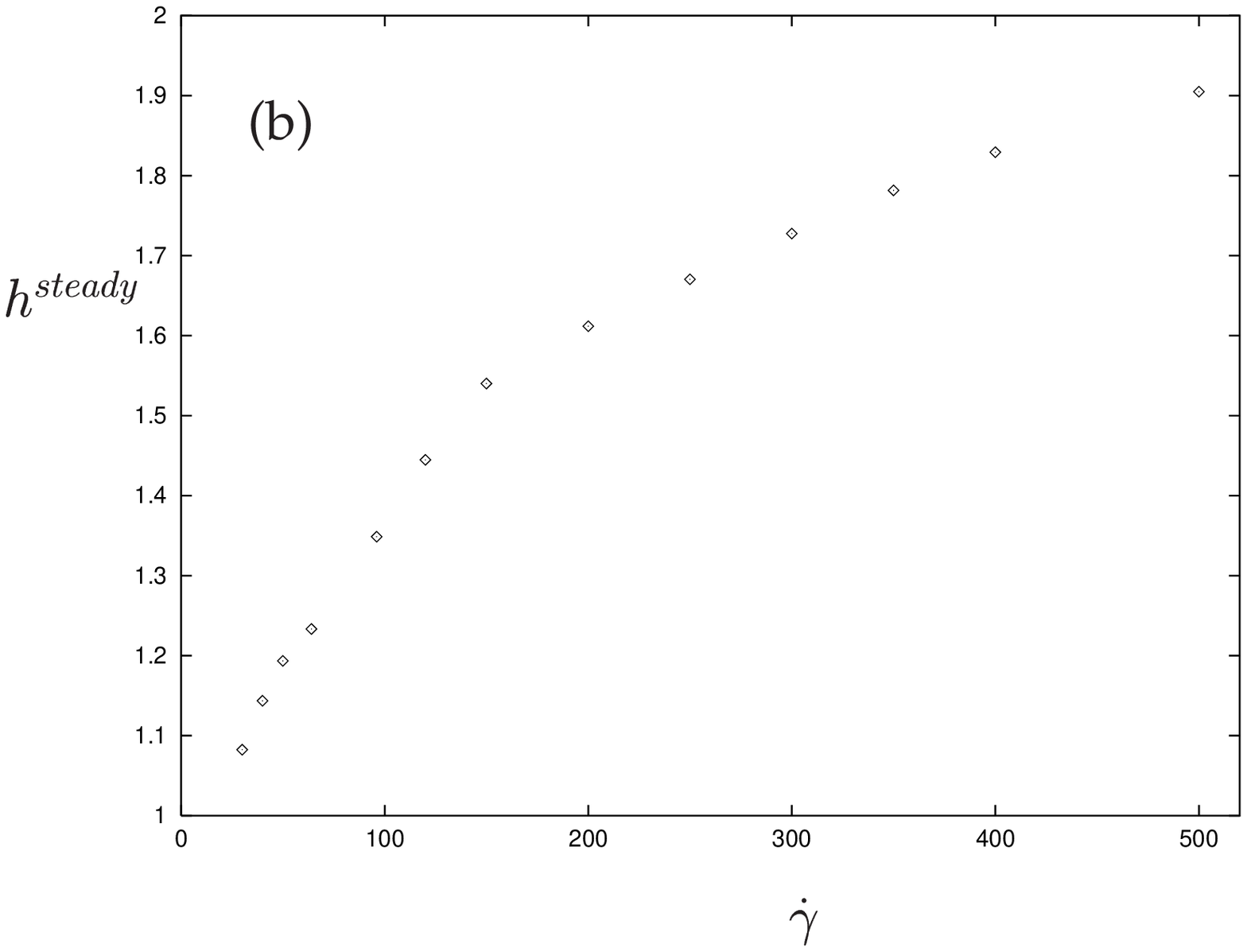}
\vspace{1cm}
\narrowtext
\caption{ 
(a) Mean vertical height of the vesicle $h$ versus $\dot{\gamma} t $ for
different shear rates. 
The shear rates, $\dot{\gamma}$, are ($a$) 64 ($b$) 96 ($c$) 150 
($d$) 200 ($e$) 300
($f$) 350 ($g$) 400 ($h$) 500. 
(b) Steady state height of the vesicle $h$ versus the shear 
rate, $\dot{\gamma} $. Here, $v = 0.95 $
and $\dot{\gamma}$ is given in dimensionless units as described 
in the text.}
\end{center}
\end{figure}

\eject

\begin{figure}
\begin{center}
\vspace{2cm}
\includegraphics[width=6in]{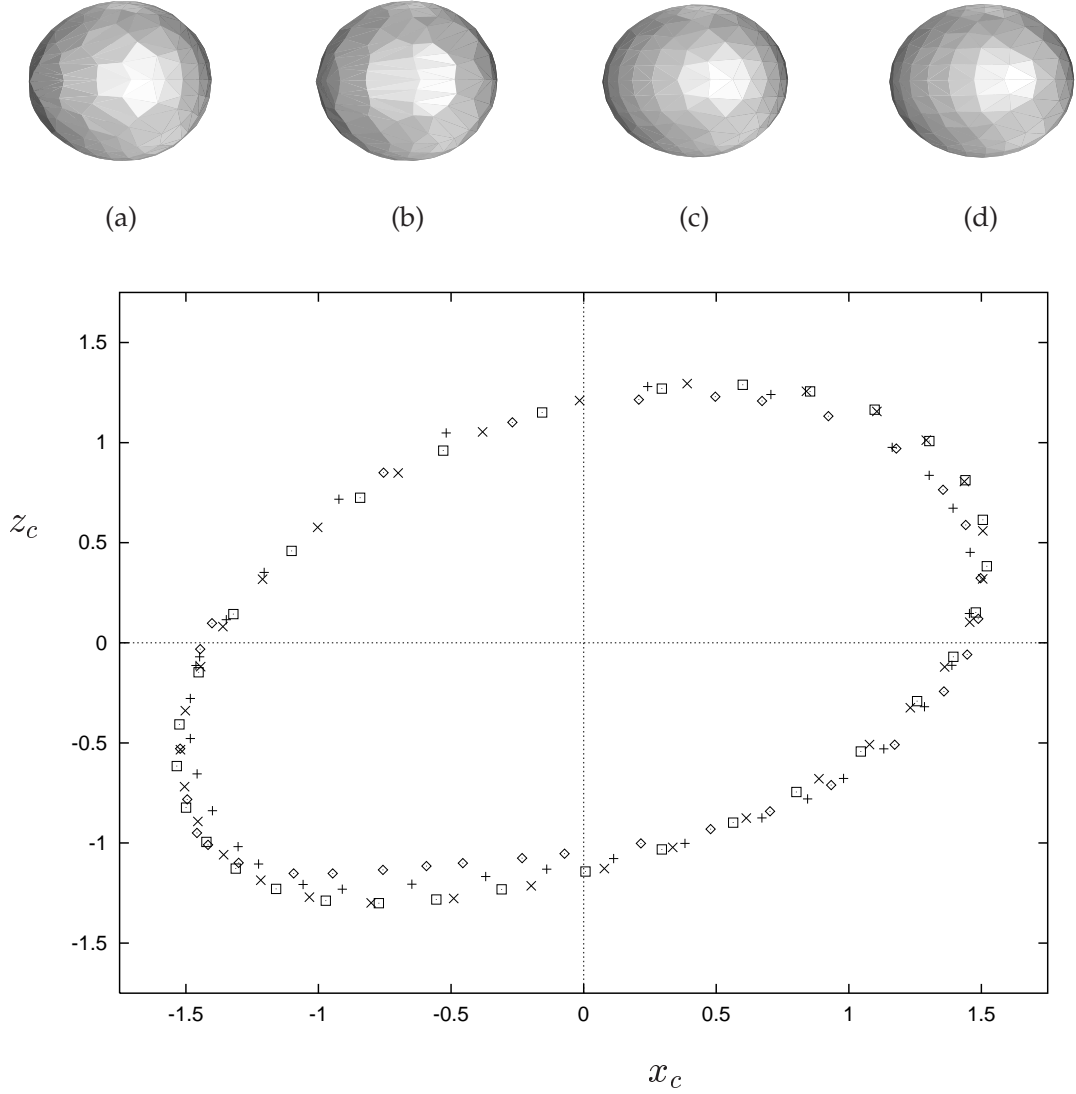}
\vspace{1cm}
\narrowtext
\caption{ The profile of a vesicle in the $x-z$ plane and the 
top view (looking down 
$z$--axis) are shown for four different shear 
rates (in dimensionless units). 
(a) $\dot{\gamma} =  64 $ ($\diamond$),
(b) $\dot{\gamma} =  96 $ ($+$), (c) $\dot{\gamma} =  350 $ ($\times$) 
and (d) $\dot{\gamma} =  400 $ ($\Box$). The profile is shown 
on axes $x_c - z_c$ such that
their mean centers coincide. Here, $v = 0.95 $ and $ g = 6.4 $.}
\end{center}
\end{figure}

\eject

\begin{figure}
\begin{center}
\vspace{2cm}
\includegraphics[width=5in]{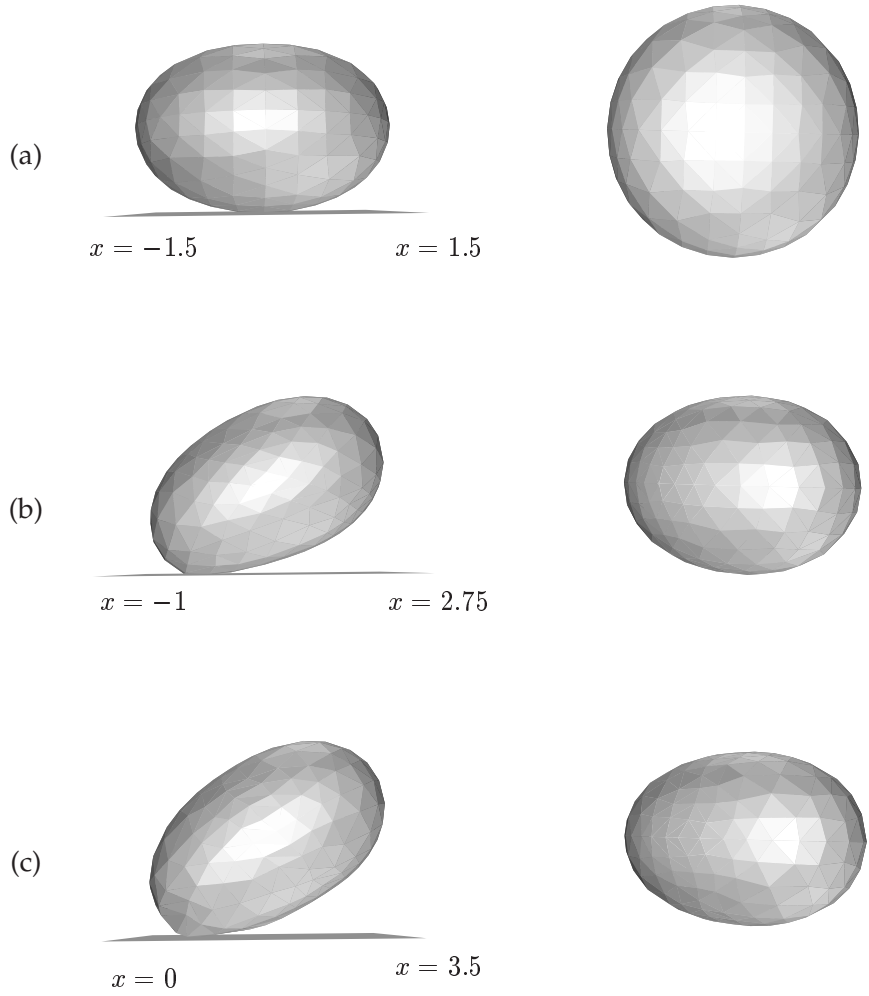}
\vspace{1cm}
\narrowtext
\caption{ The dynamical evolution of an adhering vesicle is shown.
Here, $ v = 0.95 $, $ \dot{\gamma} = 2 $ and $ W = 0.1 $ (all quantities 
scaled to be dimensionless as described in the text). The top 
and side views 
are shown on the right side and left side respectively.
(a) An adhering vesicle in equilibrium resembles an oblate 
spheroid at $t=0$. (b) In the transient stage 
the shape changes to a tilted prolate--like ellipsoid, ($t = 0.6$). 
(c) In the steady state, $ t=1.2$, the tank-treading 
vesicle is ``pinned'' to the wall and slips/rolls along the wall. The 
wall in the plane $ z = 0 $ is  
in the side view. The $x$--coordinate at the endpoints of the wall 
is shown at each instance of time. }
\end{center}
\end{figure}

\eject

\begin{figure}
\begin{center}
\vspace{2cm}
\includegraphics[width=6in,angle=0]{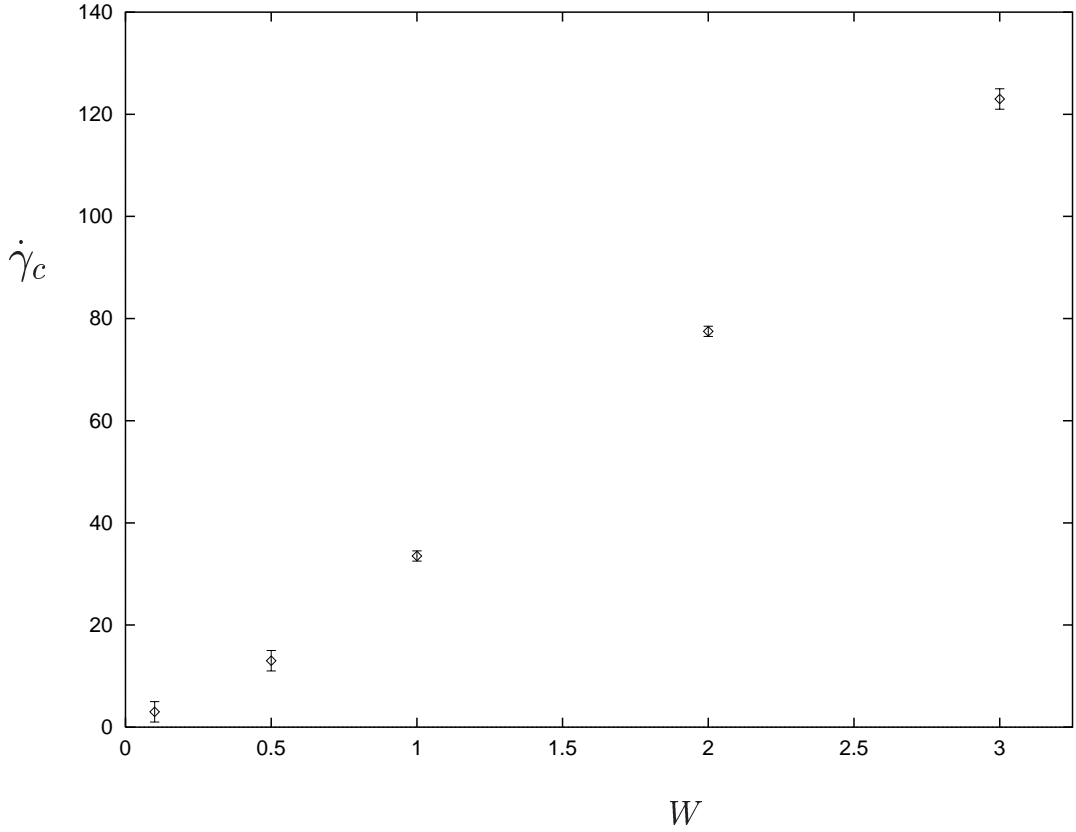}
\vspace{1cm}
\narrowtext
\caption{ The critical shear for unbinding, $ \dot{\gamma}_c$, is 
plotted against 
the adhesion strength $ W$ for $v = 0.95$, and 
$ \dot{\gamma}_c$ is linearly proportional to the
adhesion strength $ W$.
Here, $d_0 = 0.006 $. With $ \kappa = 10^{-19} J $,
$ \eta = 10^{-3} Js/m^3 $, $ R_0 = 5 \mu m $,
and with $W$ roughly $ 10^{-10}-3 \times 10^{-9} J/m^2 $, the 
critical shear rate
is roughly in the range $ 0.1-4 s^{-1} $.}
\end{center}
\end{figure}

\eject

\begin{figure}
\begin{center}
\vspace{2cm}
\includegraphics[width=6in]{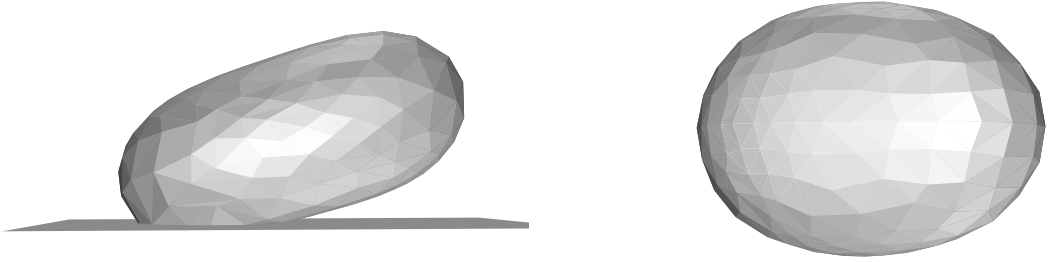}
\vspace{2cm}
\narrowtext
\caption{ An example of a ``bound'' state. 
Here, $v = 0.87 $, $ \dot{\gamma} = 2 $,
$W = 0.2$ (in dimensionless units as described in the text). The 
side and top view are shown on the left side and right side 
respectively.}
\end{center}
\end{figure}

\eject

\begin{figure}
\begin{center}
\vspace{2cm}
\includegraphics[width=5in]{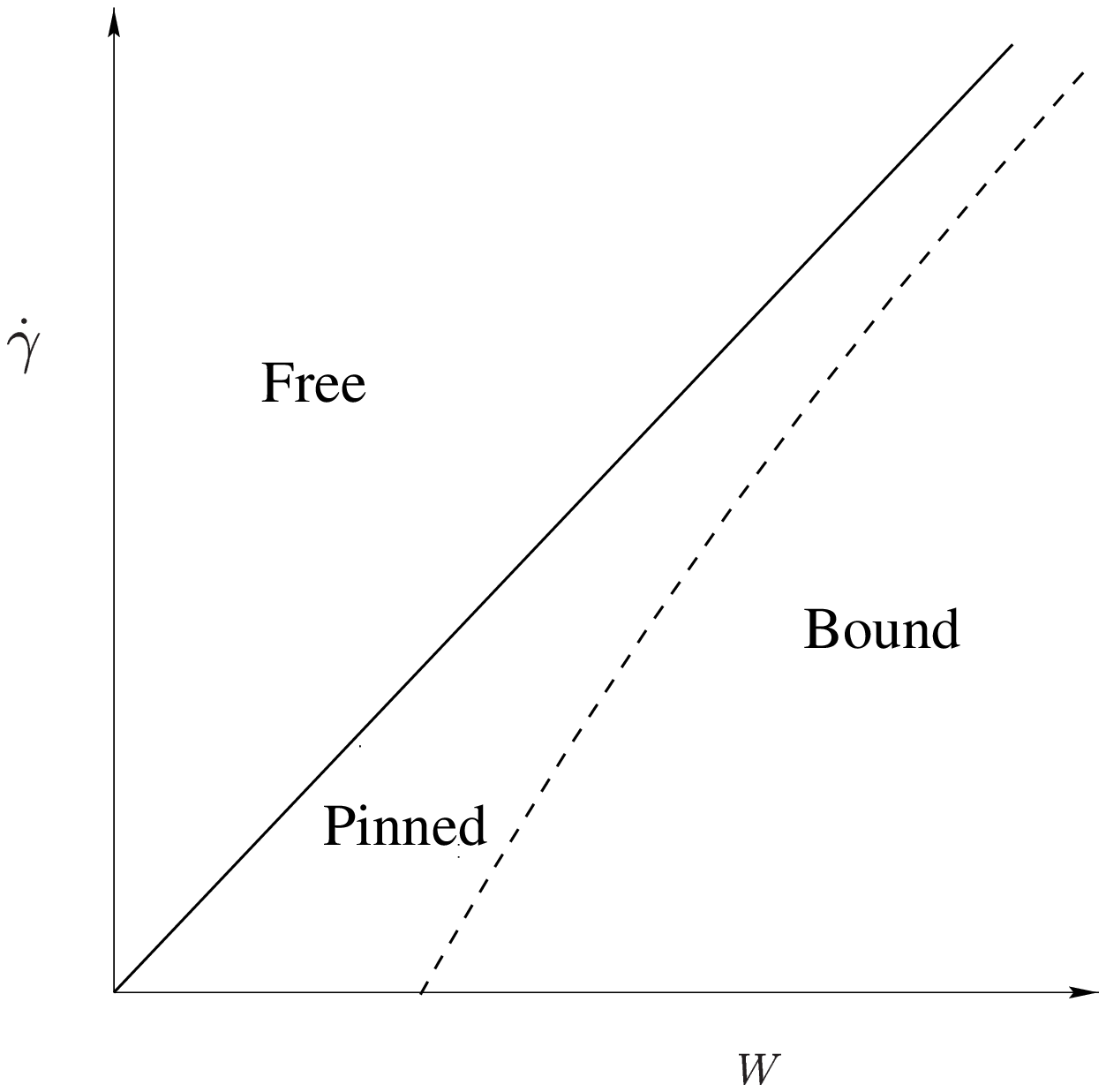}
\vspace{1cm}
\narrowtext
\caption{ Schematic phase diagram in the shear rate $\dot{\gamma}$, 
adhesion strength $W$
plane with free, pinned and bound shapes for vesicles
with arbitrary area and volume. }
\end{center}
\end{figure}


\begin{references}
\bibitem{note4} For the sake of
completeness, we note that it was pointed out \cite{olla} that
the tank-treading frequency $\omega$ and
the steady tilt $\theta$ as computed for this model show remarkable 
agreement
with a very different model, in the spheroidal limit,
which has an analytical solution \cite{keller}. This exactly
solvable model consists of a tank--treading ellipsoidal membrane with 
fixed shape and
the boundary condition on the cell surface is that the velocity be a
generic linear function of the coordinates. The analytical results are:
the tilt is given by, $ \cos 2\theta = (l_3^2 - l_1^2)/(l_3^2 + l_1^2) $, 
and
the rotational frequency $ \omega = (l_1/l_3)/(1 + (l_1/l_3)^2) $, in 
units mentioned before.
Here, $l_1$ and $l_3$ are the short axis and the long axis respectively.
\end{references}
\end{document}